\documentclass[prd,aps,preprint,showpacs]{revtex4}
\usepackage{epsfig}
\input axodraw.sty

\begin{document}

%\begin{titlepage}

%\voffset 1.5cm

%\preprint{ KIAS--P04nnn,  \hspace{1ex} hep-ph/0404029}
%

\title{\bf  Late Leptogenesis from Radiative Soft Terms }

\author{Eung Jin Chun}

\affiliation{Korea Institute for Advanced Study, 207-43
Cheongryangri 2-dong Dongdaemun-gu, Seoul 130-722, Republic of
Korea}

\date{April 2004}

\begin{abstract}
We point out that the so-called ``soft leptogenesis'' can occur at
TeV scale if the $B$ term is generated through radiative
corrections which involve two-loop diagrams with the gaugino
exchange. This mechasnism requires the non-trivial CP phase,
$\mbox{Im}(A m_{1/2}^*)\neq 0$, and can naturally explain the
observed baryon asymmetry of the universe associated with the TeV
scale seesaw mechanism. Such a low scale leptogenesis would be a
promising option in view of the tight upper limit on the reheat
temperature avoiding the gravitino problem in supergravity models.
\end{abstract}


\maketitle
%\end{titlepage}

\voffset 0cm

\noindent {\bf 1. Introduction}

The seesaw mechanism \cite{SS} provides an elegant explanation not
only for the observed neutrino masses and mixing but also for  the
baryon asymmetry of the Universe, $n_B/s \approx 10^{-10}$, under
the name of leptogenesis.  In its original suggestion \cite{FY},
it was observed that an out-of-equilibrium decay of a heavy right
handed neutrino (RHN) generates lepton asymmetry, provided CP
violating phases in the neutrino Yukawa couplings. Then, the
produced lepton asymmetry is transformed to the baryon asymmetry
through the standard model sphaleron process.    The whole scheme
can be supersymmetrized in a straightforward manner. The RHN $N$
accompanied by its scalar partner $\tilde{N}$ is extended to a
chiral superfield $\hat{N}$ which is  added to the minimal
supersymmetric standard model and allows the following
superpotential;
\begin{equation} \label{WN}
W = \mu \hat{H}_1 \hat{H}_2 + h \hat{L} \hat{H_2} \hat{N}
 + {1\over2} M \hat{N} \hat{N} \,.
\end{equation}
Here  $\hat{L}$ and $\hat{H}_{1,2}$ are the lepton and Higgs
superfields, respectively.  Now, the decays of the singlet
neutrino $N$, sneutrino $\tilde{N}$ and antisneutrino $
\tilde{N}^\dagger$ can contribute to the generation of lepton
asymmetry defined by
\begin{equation} \label{epsl}
 \epsilon \equiv { \sum_X \left[\Gamma(X \to L, \tilde{L})
- \Gamma(X \to \bar{L}, \bar{\tilde{L}}) \right] \over \sum_X
\left[ \Gamma(X \to L, \tilde{L}) + \Gamma(X \to \bar{L},
\bar{\tilde{L}}) \right]}
\end{equation}
where $X$ denotes $N$, $\tilde{N}$ or $\tilde{N}^\dagger$. In the
conventional ``supersymmetric leptogeneis'' \cite{CRV,Plu}, each
particle $X$ contributes to  the CP asymmetric quantity
(\ref{epsl}) in the same way as described before. Namely, the
required CP violation arises from the interference between the
tree-level and one-loop diagrams with non-trivial CP phases in the
Yukawa coupling $h$.

Recently, the importance of supersymmetry breaking effect in
supergravity models has been realized.  According to the original
observations of Refs.~\cite{GKNR,AGR}, an enhanced contribution to
leptogenesis may occur through a tiny
$\tilde{N}$--$\tilde{N}^\dagger$ mixing requiring a non-trivial CP
phase, $\mbox{Im}(A B^*)\neq 0$ where $A,B$ are dimension-one soft
parameters for the scalar RHN. The resulting lepton asymmetry in
the so-called ``soft leptogenesis'' is given by
\begin{equation}
\epsilon \approx {\mbox{Im} (A B^*) \over \Gamma^2 +  |B|^2}\,
{\Gamma \over M}\, \Delta_{BF}
\end{equation}
where $\Delta_{BF}$ counts for the supersymmetry breaking effect
at a finite temperature \cite{CRRV,GKNR,AGR}.

An important  constraint on the leptogenesis  scenario in the
supergravity models  comes from the gravitino problem, which
requires the reheat temperature (and thus the RHN mass $M$) to be
low enough in order not to overproduce gravitinos \cite{EKN}.
Otherwise, the late decay of gravitinos upsets the standard
prediction of the nucleosynthesis on  the primordial abundance of
the light elements.  Recent analysis has  put  a very strong upper
bound on the reheat temperature, $T_R \lesssim
(2\times10^6-2\times10^7)$ GeV for the gravitino mass in the range
of $m_{3/2} =(10^2-10^3)$ GeV \cite{moroi}.  As is well-known, the
supersmmetric leptogenesis cannot work with  such a low reheat
temperature since the following bound has to be fulfilled
\cite{Mbound}:
\begin{equation} \label{Mlower}
 M \gtrsim 10^9\, \mbox{GeV} \left( \epsilon \over 10^{-7} \right)
          \left( 0.05\,\mbox{eV} \over m_{\nu_3} \right)
\end{equation}
where $M$ is taken to be the lowest seesaw scale.   The similar
constraint also applies to the soft leptogenesis where the typical
values of the  soft terms are assumed; $A \sim B \sim m_{3/2}$. It
is however possible to have much smaller seesaw scale if the $B$
parameter is arranged to satisfy \cite{GKNR,AGR}:
\begin{equation}
 B \approx 10^{-3}\mbox{ GeV} \left( M \over 10^7 \mbox{ GeV}
 \right)^2 \,.
\end{equation}

In this paper, we show that the soft leptogenesis works
successfully for the TeV scale seesaw mechanism  provided the
initial condition of the soft term:
\begin{equation} B=0
\,.\end{equation}
 This mechanism involves the two-loop diagrams
carrying gauginos in the loop, and the corresponding CP phases
from the soft supersymmetry breaking $A$ term and the gaugino mass
$m_{1/2}$ \cite{tevL}. While it is difficult to arrange a
hierarchically small value of $B \ll m_{3/2}$ in general
supergravity models, the initial condition $B=0$ may arise
naturally in the no-scale type models. We also work out a concrete
model in which  such a feature is realized and  the TeV seesaw
scale $M$ \cite{tevM} can be understood in connection with the
Higgsino mass parameter $\mu$.

\medskip

\noindent {\bf  2. CP phases from soft supersymmetry breaking}

Let us first make a general description of CP violation coming
from soft supersymmetry breaking which leads to a non-trivial
lepton asymmetry. The scalar potential from soft supersymmetry
breaking can be written as
\begin{equation} \label{Vsoft}
V_{soft} = m_0^2 |\phi|^2 + \left[ B\mu H_1 H_2 + A h \tilde{L}
H_2 \tilde{N} + {1\over2} B M \tilde{N}\tilde{N} + h.c.\right]
\end{equation}
where $\phi$ represents any scalar field, and $m_0$, $A$ and $B$
are dimension-one soft masses. Simply extending the standard
argument of Ref.~\cite{DS}, we can  consider mass parameters as
spurion fields and then find that our Lagrangian possesses the
$U(1)_R$ and $U(1)_{PQ}$ symmetry defined by the following $R$ and
$PQ$ charges:
\begin{equation}
\begin{array}{rrrrrrrrrr}%{cccccccccc}
   & m_0^2& A & B & m_{1/2}& \mu & N & L& H_1 & H_2 \cr
R & 0 & -2 & -2 & -2 & 0 &1 & 0 & 1 & 1 \cr PQ & 0 & 0 & 0 & 0 &
-2 & 0 & -1 & 1 & 1 \cr
\end{array}
\end{equation}
 Here $m_{1/2}$ is the gaugino mass. Since physical observables
should be neutral under such symmetries, the lepton asymmetry
$\epsilon$ (\ref{epsl}) can only depend on one of the following
combinations;
\begin{equation}
\quad AB^*,\quad A m^*_{1/2},\quad B m^*_{1/2} (\mu\mu^*)
\end{equation}
which shows that  we have only two independent phases as usual
\cite{DS}. Note that the leptogenesis from the
$\tilde{N}$--$\tilde{N}^\dagger$ mixing requires the non-trivial
CP phase of the first type $AB^*$ \cite{GKNR,AGR}.    As a simple
collorrary of our argument, one sees that the  supersymmery
breaking contribution to leptogenesis occurs at the order of
$\tilde{m}^2$  at most.   In the following, we will discuss the
gaugino contribution with a non-trivial phase in the combination
$A m^*_{1/2}$.

\medskip

\noindent {\bf 3. Gaugino-loop  contribution to soft
leptogenesis}

From the interaction terms in Eqs.~(\ref{WN}) and (\ref{Vsoft}),
one obtains the following lepton number producing processes of the
singlet sneutrinos $\tilde{N}$ and $\tilde{N}^\dagger$:
$\tilde{N}^\dagger \to L \tilde{H}_2,\, \tilde{L} H_2,\, \tilde{L}
H_1^\dagger$ and $\tilde{N}\to \tilde{L} H_2$ which have the
coupling $h,\, h A^*,\, h\mu$ and $hM$, respectively. There are
also the usual singlet neutrino decay; $N\to L H_2$ and $\tilde{L}
\tilde{H}_2$ with the coupling $h$. For the convenience of our
discussion, we take $h$ and $M$ to be real without a loss of
generality.

Assuming the boundary condition $B=0$ imposed in a certain
supersymmetry breaking mechanism, the $B$ term received two
important radiative corrections.  One comes from the one-loop
diagram having $\tilde{L}$ and $H_2$ in the loop with couplings
$hM$ and $hA$, and the other is the two-loop diagram drawn in
Fig.~1.

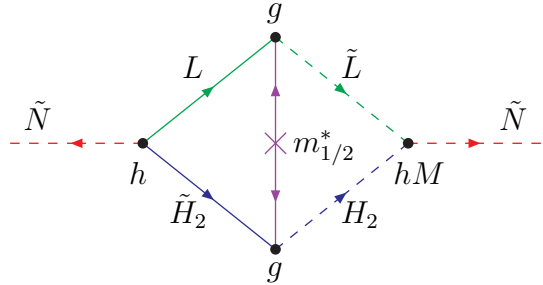
\begin{center}
\begin{picture}(200,130)(0,0)
\SetColor{Red} %N line
\DashArrowLine(50,50)(0,50){4} \DashArrowLine(150,50)(200,50){4}
\SetColor{Green}  %lepton line
\ArrowLine(50,50)(100,90) \DashArrowLine(100,90)(150,50){4}
\SetColor{Blue}  %higgs line
\ArrowLine(50,50)(100,10) \DashArrowLine(100,10)(150,50){4}
\SetColor{Purple} %gaugino line
\ArrowLine(100,50)(100,10) \ArrowLine(100,50)(100,90)
\Line(96,54)(104,46) \Line(96,46)(104,54)
\SetColor{Black} %vertices
\Vertex(50,50){2} \Vertex(100,90){2} \Vertex(100,10){2}
\Vertex(150,50){2} \put(45,35){$h$} \put(145,35){$hM$}
\put(97,0){$g$} \put(97,98){$g$} \put(5,55){$\tilde{N}$}
\put(185,55){$\tilde{N}$} \put(125,20){$H_2$}
\put(125,75){$\tilde{L}$} \put(60,20){$\tilde{H}_2$}
\put(65,75){$L$} \put(107,47){$m_{1/2}^*$}
%\SetColor{Black} % cuts
%\PhotonArc(100,0)(25,70,150){1}{3}
%\PhotonArc(100,100)(25,210,290){1}{3}
\end{picture} \\ {\sl Fig.~1.  Two-loop diagram with gaugino leading to
non-vanishing $B$ parameter and thereby the desired soft
letogenesis at TeV scale. }
\end{center}

From these diagrams, one finds
\begin{equation}
 B \approx \left[ {1 \over \pi} A   + {\alpha_2 \over
 4\pi^2} m_{1/2} \right] {\Gamma\over M} \ll \Gamma
\end{equation}
where $\alpha_2$ is the gauge coupling constant $g^2/4\pi$. While
the first term does not contribute to the lepton asymmetry
$\epsilon$ (3), the second term gives
\begin{equation} \label{epsm}
\epsilon \approx {\alpha_2\over 4\pi^2} {\mbox{Im}(Am^*_{1/2})
\over M^2 } \Delta_{BF} \,.
\end{equation}
Taking $\mbox{Im}(A m_{1/2}^*) = 10^4$ GeV$^2$, $M=1$ TeV and
$\Delta_{BF}=0.1$, we obtain the right range of the lepton
asymmetry $\epsilon \sim 10^{-6}$.  Note that we have ignored the
lepton asymmetry coming from the supersymmetry breaking vertex
with the $hA$ coupling.  This is a good approximation in the
regime of $|A| \ll M$ where one finds
\begin{equation} \label{epsn}
 \epsilon \sim {\alpha_2\over 4\pi^2} {\mbox{Im}(Am^*_{1/2})
\over M^2 } {|A|^4 \over M^4}\,.
\end{equation}
This, of course, becomes important for $A\sim M$ and the above
order of magnitude estimation of the lepton asymmetry still holds.
Let us remark that the lepton asymmetry (\ref{epsm}, \ref{epsn})
is independent of $h^2$ or the decay rate $\Gamma$, and thus
independent of the neutrino mass. That is, the usual
out-of-equilibrium condition $\Gamma < H$ has no direct impact on
the lepton asymmetry apart from its effect on the quantity
$\Delta_{BF}$.

\medskip

\noindent {\bf 4. A model}

In this section, we consider the possible origin of the TeV seesaw
scale $M$ and the initial condition $B=0$.   For this purpose, we
invoke the model \cite{LPQ} where the lepton number symmetry is
promoted to the Peccei-Quinn  ($PQ$) symmetry  which is introduced
to solve the strong CP problem \cite{kim1}.  The original
motivation was to relate two intermediate scales, the seesaw scale
$M$ and the $PQ$ symmetry breaking scale $f_{PQ}$, both of which
are around $10^{11}$ GeV \cite{LPQ}.  In this scheme, the Higgsino
mass parameter $\mu$ is generated from the non-renormalizable term
which is dictated by the non-trivial $PQ$ charge assignment to the
Higgs bilinear $H_1 H_2$ \cite{muterm}. Taking now a different
point of view, it is conceivable to generate two TeV scale
parameters $\mu$ and $M$  from the $PQ$ symmetry breaking.  To be
more specific,  let us introduce the singlet fields ($P$, $Q$ and
$S$) which are responsible for the spontaneous breaking of the
$PQ$ symmetry as is done usually. To achieve the desired pattern
for the soft terms, we also introduce the supersymmetry breaking
field $Z$ in the hidden sector with $\langle Z \rangle \sim M_P$,
$\langle F_Z \rangle \sim m_{3/2} M_P$.  Assigning the  $PQ$ and
$R$ charges to the fields as follows;
\begin{equation}
\begin{array}{rrrrrrrrrr}%{cccccccccc}
   &  L &  E^c &  N &  H_1 &  H_2 &  P &  Q &  S & Z        \cr
   PQ
& 0 &  -1  &  -1 &  1  &   1  &  -1 & 1 &  0  & 0           \cr
 R
& -1& 1 &   1 &  1  &   1  &   0 & 0 &   2  & 1   \cr
\end{array}
\end{equation}
one obtains the allowed superpotential written as
\begin{equation}
 W = h_e e^{Z/M_P} L H_1 E^c + h_N e^{Z/M_P} L H_2 N
 + h_\mu {P^2\over M_P} H_1 H_2 + h_M {Q^2\over M_P} N N
 + h_S S(PQ -f_{PQ}^2) \,.
\end{equation}
Here,  we introduced one intermediate scale $f_{PQ} \sim
\sqrt{m_{3/2} M_P}$  which might be related to the supersymmetry
breaking scale $\langle F_Z \rangle$.  Note that the $R$ symmetry
is non-linearly realized by the supersymmetry breaking field $Z$
which behaves like a dilaton field in superstring theories.
 Now, a crucial ingredient is to {\it assume no-scale supergravity} with
$K= -3\log( Z + Z^*- \phi\phi^*)$ where $\phi$ denotes the
observable (and $PQ$) sector fields.  In the general supergravity
models, the $B$ term receives the contribution of the order
$m_{3/2}$ coming from various terms in the supergravity scalar
potential \cite{brignole}.  The condition $B=0$ could be made by
arranging cancellation among various terms, which is however
unlikely to occur. As is well-known, the no-scale supergravity
provides a restrictive prediction on the soft terms
\cite{brignole}, which is also useful for our purpose.  With the
above superpotential (11), one finds the universal $A$ and $B$
terms as follows;
\begin{equation}
 A = {\langle F_Z \rangle \over M_P} \sim m_{3/2}
  \quad\mbox{and}\quad
 B=0
\end{equation}
together with $\mu \sim M\sim f_{PQ}^2/M_P\sim m_{3/2}$ as alluded
before.

Here, let us remark that the mechanism of dynamical relaxation for
$B=0$ was considered in Ref.~\cite{masahiro}, which is advocated
to resolve the associated supersymmetric CP problem. Such a
resolution also works in  our model as both $B$ terms for the
Higgs  and scalar RHN sectors vanish.

\medskip

\noindent {\bf 5. Conclusion}

Leptogenesis is a promising way of generating the baryon asymmetry
of the Universe which is linked with the origin of neutrino masses
and mixing.  Besides the widely discussed supersymmeric
contribution,  leptogenesis may receive an important contribution
from supersymmetry breaking. In this paper, we point out that soft
supersymmetry breaking involving CP violation in the gaugino
sector, more precisely Im($Am_{1/2}^*$),  can be the source of a
successful leptogenesis if the $B=0$ boundary condition of the
soft parameters is realized. This turns out to work for the TeV
scale  seesaw mechanism.  Our analysis implies that leptogenesis
can be operative  even with  a fairly low seesaw scale which is
compatible with the low reheat temperature inflation which evades
the gravitino problem. Furthermore, it will be interesting to see
if certain CP violating phenomena can be observed in future
collider experiments associated with the above ``leptogensis
phase''.

\medskip

{\bf Acknowledgments:} The author would like to thank Sacha
Davidson and Yosef Nir for valuable comments and criticisms. He is
also grateful to IPPP for its hospitality during his visit. His
work was supported by the Korea Research Foundation Grant,
KRF-2002-070-C00022.

\end{document}